%% file: main.tex
\pdfoutput=1

\documentclass[sigconf, nonacm]{acmart}

\AtBeginDocument{%
  \providecommand\BibTeX{{%
    \normalfont B\kern-0.5em{\scshape i\kern-0.25em b}\kern-0.8em\TeX}}}

\setcopyright{acmcopyright}
\copyrightyear{2018}
\acmYear{2022}
\acmDOI{XXXXXXX.XXXXXXX}

\acmConference[ARES '22]{Make sure to enter the correct
  conference title from your rights confirmation email}{August 23--26,
  2022}{Vienna, Austria}
\acmPrice{15.00}
\acmISBN{978-1-4503-XXXX-X/18/06}

\usepackage{graphicx}

\usepackage{lineno}

\usepackage{booktabs} 
\usepackage{algorithm}
\usepackage{algpseudocode}
\usepackage{xcolor}
\usepackage{footnote}
\usepackage[nolist]{acronym}
\usepackage{listings}
\usepackage{caption}
\usepackage{subcaption}
\newcommand\todo[1]{\textcolor{red}{#1}}
\usepackage{array}
\usepackage{tikz}
\usepackage{pgfplots}
\usepackage{comment}
\usepackage{enumitem}
\usepackage{placeins}
\usepackage{xfrac}
\usepackage{blindtext,rotating}
\pgfplotsset{width=6.6cm,compat=1.15}

\usepackage{tabu}
\usepackage{url}

\DeclareCaptionFont{white}{ \color{black} }
\DeclareCaptionFormat{listing}{
  \colorbox[cmyk]{0.43, 0.35, 0.35,0.01 }{
    \parbox{81mm}{\hspace{15pt}#1#2#3}
  }
}

\hypersetup{
	colorlinks=true,
   	linkcolor=blue,
   	citecolor=blue,
    filecolor=blue,
   	urlcolor=blue}

\newdimen\NetTableWidth

\begin{document}



\title{A Reverse Engineering Education Needs Analysis Survey}




\author{Charles~R.~Barone~IV}
\affiliation{
    \institution{University of New Haven}
    \country{United States of America}
    }
\email{cbaro4@unh.newhaven.edu}

\author{Robert~Serafin}
\affiliation{
    \institution{University of New Haven}
    \country{United States of America}
    }
\email{rsera1@unh.newhaven.edu}

\author[newhaven]{Ilya~Shavrov}
\affiliation{
    \institution{University of New Haven}
    \country{United States of America}
    }
\email{ishav1@unh.newhaven.edu}

\author{Ibrahim Baggili}
\affiliation{
    \institution{Louisiana State University \\ Center for Computation \& Technology}
    \country{United States of America}
    }
\email{ibaggili@lsu.edu}

\author{Aisha~Ali-Gombe}
\affiliation{
    \institution{Louisiana State University \\ Center for Computation \& Technology}
    \country{United States of America}
    }
\email{aaligombe@lsu.edu}

\author{Golden~G.~Richard~III}
\affiliation{
    \institution{Louisiana State University \\ Center for Computation \& Technology}
    \country{United States of America}
    }
\email{golden@cct.lsu.edu}

\author{Andrew Case}
\affiliation{
    \institution{Louisiana State University \\ Center for Computation \& Technology}
    \country{United States of America}
    }
\email{acase8@lsu.edu}

\begin{abstract}
This paper presents the results of a needs analysis survey for \ac{RE}. The need for reverse engineers in digital forensics, continues to grow as malware analysis becomes more complicated. The survey was created to investigate tools used in the cybersecurity industry, the methods for teaching \ac{RE} and educational resources related to \ac{RE}. Ninety-three (n=93) people responded to our 58 question survey. Participants did not respond to all survey questions as they were optional. The data showed that the majority of \sfrac{24}{71} (33.8\%) responses either strongly agreed and \sfrac{22}{71} (30.99\%) of responses somewhat agreed that there is a shortage in \ac{RE} resources. Furthermore, a majority of \sfrac{17}{72} (23.61\%) responses indicated that they strongly disagree and that \sfrac{27}{72} (37.5\%) somewhat disagree  to the statement that graduates are leaving college with adequate \ac{RE} knowledge. When asked if there is a shortage of adequate \ac{RE} candidates, the majority of \sfrac{33}{71} (46.48\%) responses strongly agreed and \sfrac{20}{71} (28.17\%) somewhat agreed. In order to determine if this was a result of the tools at their disposal, a series of questions in regards to the two most popular \ac{RE} tools were also asked. In response to a question about the efficiency of the \ac{IDA} \ac{GUI}, majority of \sfrac{16}{50} (32\%) respondents strongly agreed and \sfrac{27}{50} (54\%) somewhat agreed. Similarly, in response to the same question, but in regards to the Ghidra \ac{GUI}, majority of \sfrac{14}{32} (43.75\%) respondents somewhat agreed. The results show that a significant number of respondents were self-taught and weren't educated via a university, college, or employer in \ac{RE}. It was concluded that \ac{RE} is still a growing field, with few experts and a lack of proficiency, but not necessarily impeded by a lack of adequate tools. Our paper provides an overview of the current state of the \ac{RE} community to be used for the future development of the \ac{RE} domain.

\end{abstract}  



\maketitle 


\input{sections/acronyms.tex}
\input{sections/sec_1.tex}
\input{sections/sec_2.tex}
\input{sections/sec_3.tex}

\input{sections/sec_4.tex}

\input{sections/sec_5.tex}
\input{sections/sec_6.tex}

\bibliographystyle{agsm}

\bibliography{MyCite.bib}

\input{sections/appendix_a.tex}
\input{sections/appendix_b.tex}

\input{sections/appendix_c.tex}
\input{sections/appendix_d.tex}
\input{sections/appendix_e.tex}
\input{sections/appendix_f.tex}

\end{document}

%% file: sections/acronyms.tex
\begin{acronym}
\acro{CISA}{Cybersecurity and Infrastructure Security Agency}
\acro{NSA}{National Security Agency}
\acro{RE}{Reverse Engineering}
\acro{IDA}{Interactive Disassembler}
\acro{GDB}{GNU Debugger}
\acro{UI}{User Interface}
\acro{OSS}{Open Source Software}
\acro{CTF}{Capture The Flag}
\acro{DMCA}{Digital Millennium Copyright Act}
\acro{GUI}{Graphical User Interface}
\acro{EULA}{End-User License Agreement}
\acro{IRB}{Institutional Review Board}
\acro{CPU}{Central Processing Unit}
\acro{API}{Application Programming Interface}
\acro{ARM}{Advanced RISC Machines}
\acro{WinDbg}{Windows Debugger}
\acro{CAE-CO}{Centers of Academic Excellence in Cyber Operations}
\end{acronym}

%% file: sections/sec_1.tex
\section{Introduction}
\label{sec:introduction}

The origins of \ac{RE} can be traced back to hardware, where physical objects were deconstructed in order to determine design logic as well as functionalities. While hardware \ac{RE} is still prevalent today, there is a significant shift in focus to \ac{RE} software, especially within the security community. A formal definition of software \ac{RE} was presented by \citep{chikofsky1990reverse} in 1990, as ``...the process of analyzing a subject system to identify the system's components and their relationships and create a representation of the system in another form or at a higher level of abstraction.'' Over the years, software \ac{RE} has gained significant momentum, particularly in the areas of program analysis and re-engineering and transformation \citep{conti2008visual, canfora2007new}. Various \ac{RE} toolkits were created that support a number of  different architectures, assist with design recovery, and allow visualization of extracted data.

Although these advancements are significant, with the increasing complexity of cyberattacks and proliferation of highly stealthy and obfuscated malware, the need for software \ac{RE} in the digital forensics and incident response continues to grow. Today, \ac{RE} plays a critical role in the deep understanding and attempted mitigation of malicious software \citep{richard2009highly}, which \citep{luciano2018digital} stated is imperative to digital forensics. Major companies, such as Target, Equifax, and Apple, have been compromised and experienced the leakage of user data for millions of users. Additionally, cities in the United States have had to pay a significant amount of money to regain access to their systems after sophisticated ransomware attacks, as seen in  Florida towns \citep{norwood_2019}. The Cybersecurity and Infrastructure Security Agency \ac{CISA} director Christopher Krebs has gone on to state that ransomware attacks are a national crisis \citep{seals_2020}. Finally, the United States government's response to combating cyber attacks further demonstrates this urgency. The U.S. Fiscal Year Budget for 2020 has allocated 18.4 billion dollars for cybersecurity purposes to a number of agencies within the federal government \citep{better_america, cybersecurity_funding_2019}. 

While it is important to highlight technical advancements in \ac{RE} and to continue to build powerful tools for \ac{RE}, it is also critical that we try to understand outstanding issues that plague this field so that we can focus our energy on effective solutions.  \citep{selfridge1993challenges} points out some fundamental problems of \ac{RE}, such as using example programs, understanding economic impact, and inter-researcher communication. It's also clear that there are significant barriers to entry for students.  These include the availability of appropriate training, the need to use complex or expensive software packages, and the lack of prerequisite knowledge.  The latter is a serious issue, as the skills and knowledge most helpful in \ac{RE}, including advanced systems programming, operating systems internals, assembler language, compilers, and more, are increasingly marginalized in academic programs \citep{richard2009highly}.

In this paper, we present the results of a detailed needs survey for \ac{RE}.  Our survey seeks to explore the general need for \ac{RE} in the cybersecurity and digital forensics communities and how \ac{RE} is taught in universities and learned via on-the-job training. Similar needs analysis surveys have been conducted in the literature for other fields of cybersecurity, such as cyber forensics \citep{rogers2004future, harichandran2016cyber}. We undertook this effort because it is crucial for the growth of \ac{RE} as a whole to reflect on the current standing of the field and its challenges. Our contributions are as follows:
\begin{itemize}[nosep]
  \item We present the primary account of a needs analysis survey for \ac{RE}. Our 58 question survey gained the feedback from (n=93) respondents, for the analysis of \ac{RE} needs.
  \item We present primary data on how people learn, teach and work in \ac{RE} while also explaining what challenges they face during that process. 
  \item Our analysis provides insight for advancing the \ac{RE} in education and tools.
\end{itemize}

The paper is divided into several sections. First, we discuss background information on \ac{RE} in Section \ref{sec:background}. In Section \ref{sec:methodology}, we share the methodology and survey design. The results are presented in Section \ref{sec:results}. Section \ref{sec:limitations} describes the limitations of our work, and Section \ref{sec:related work} shares related work. Lastly, Section \ref{sec:discussion/conclusions} discusses and concludes our work. 

%% file: sections/sec_2.tex
\section{Background}
\label{sec:background}
In recent years, software \ac{RE} has played a pivotal role in cybersecurity and digital forensics \citep{akbanov2018static,burji2010malware,rahimian2013reverse}.  
There are basically two major static analysis techniques for software \ac{RE} - disassembly and decompilation. Disassemblers are tools that generate low-level assembly code from a compiled binary for further scrutiny. On the other hand, decompilation is the process of translating an executable into  high-level source code. These RE techniques are widely used in various computing fields ranging from software development to malware and vulnerability analysis. In software development, \ac{RE} provides a mechanism for recovering and recreating a higher-level abstraction of legacy software for maintenance and patching even where the original source code is not available. In addition, these \ac{RE} techniques can be applied for the recovery of source code, architecture and design discovery \citep{canfora2007new}, and the integrity validation of cryptographic algorithms \citep{eilam2011reversing}. In the malware analysis domain, \ac{RE} is applied to identify the malicious payload, obfuscation, and other functionalities. 

\subsection{Current Tools}
A variety of tools are used to reverse engineer applications. Generally, these tools can be separated into two different categories: static analysis and dynamic analysis. Static analysis tools examine the application without executing it while dynamic analysis tools examine the application during execution. These categories are not mutually exclusive--some tools provide both static and dynamic analysis functionality. While open source tool development for \ac{RE} is growing, reverse engineers typically use a mixture of open and closed source tools.  

\subsubsection{Static Analysis Tools}
\ac{IDA} by Hex Rays is a disassembler and decompiler that is interactive, programmable and extensible. \ac{IDA} is hosted on Windows, Linux, or Mac OS X \citep{ida}. It was created by Ilfak Guilfanov, who later founded Hex Rays. Ghidra is a suite of tools developed for software \ac{RE} created by the \ac{NSA}'s Research Directorate in support of the cybersecurity mission \citep{ghidra}. It was publicly released by the \ac{NSA} on March 2019, with the source code being released on Github \citep{nationalsecurityagency_2020}.  Radare2 is an open source \ac{RE} tool suite that initially provided only static analysis capabilities, but now also provides an integrated low-level debugger \citep{radare2}. 

\subsubsection{Dynamic Analysis Tools}
There are a variety of publicly available tools that can be used to perform dynamic analysis of an executable. The application will typically execute in a sandbox or virtual environment (e.g., Virtualbox, VMWare Workstation, Microsoft Hyper-V) to avoid contamination of the analysis machine as the user monitors application behavior. There are also web-based tools that allow users to upload files to be scanned in order to determine if they are malicious. Some examples are Virustotal and Hybrid Analysis. x64dbg is an open-source binary debugger for Windows, aimed at malware analysis and \ac{RE} of executables for which source code is not available \citep{x64dbg}.  Immunity Debugger builds on a solid \ac{UI} with function graphing, the industry's first heap analysis tool built specifically for heap creation, and a large and well supported Python \ac{API} for easy extensibility \citep{Immunitydbg}. 

The \ac{WinDbg} can be used to debug kernel-mode and user-mode code, analyze crash dumps, and examine the \ac{CPU} registers while the code executes \citep{windbg}. OllyDbg is a debugger that operates at the assembler level on Windows and focuses on binary code analysis. It is free and open source, which makes it a very popular debugging solution \citep{ollydbg}. Cheat Engine is a tool created for use primarily with single player video games and allows a user to scan for and change values of variables that control game behavior. It also includes a debugger, disassembler, assembler, and system inspection tools which can be of use to someone performing analysis of other types of applications \citep{cheatengine}.

%% file: sections/sec_3.tex
\section{Methodology}
\label{sec:methodology}
Given the different flavors of \ac{RE} tools and the different ways in which practitioners use the tools for analysis and training, it is clear that there is no harmonized approach to \ac{RE} within the security and forensics communities. Thus, to understand the driving factors and motivation for adopting specific tools or collection of tools by practitioners and educators, we conduct a need analysis survey. The survey was created and taken  on  Qualtrics. We began by obtaining a category two exemption from the \ac{IRB} at the BLINDED FOR REVIEW
(this meant that the survey did not record participant identification information or behavior, and posed no risk or harm to subjects not encountered in every day life). The recorded responses from the survey were then analyzed using various exploratory analysis methods.

\subsection{Survey    Design \& Data Analysis}
\label{sec:surveydesign}
The survey was mainly geared towards individuals who were professionals or educators in the cybersecurity and digital forensics fields. It consisted of 58 questions with the following composition:
\begin{itemize}[nosep]                      
    \item 10 Likert Scale               
    \item 21 Multiple Choice            
    \item 15 Multiple Choice (Checkbox) 
    \item  9 Free Response              
    \item  1 Slider                   
\end{itemize}
The survey questions covered the following: demographics, teaching \ac{RE}, learning \ac{RE}, \ac{RE} Tools and platforms, \ac{IDA} versus Ghidra, potential for web based \ac{RE} platforms, and legal challenges in adopting and usage of \ac{RE}. Collected responses were analyzed using exploratory data analysis and visualization techniques.  




%% file: sections/sec_4.tex
\section{Results}
Ninety-three participants were recorded to have accessed the survey. All (n=93) participants consented to take the survey, three did not continue the survey beyond the definitions and seven did not record any responses beyond the first question. Not all participants completed every question in the survey. This is discussed further in the Limitations Section \ref{sec:limitations} of the paper.
\label{sec:results}
\subsection{Demographics}
The demographic results as shown in Appendix \hyperref[sec:appendixa]{A}, Table \hyperref[table:a1]{A.1} shows that the majority of the responses \sfrac{56}{85} (65.88\%) were white, \sfrac{69}{85} (81.18\%) were males and \sfrac{30}{85} (35.29\%) within the age range of 30-39. The majority of responses \sfrac{24}{85} (28.24\%)  indicates a Masters degree as participants highest level of education. Most of the responses indicate degrees related to the field of technology with a majority of in Computer Science \sfrac{26}{83} (31.33\%).
 
Interest in \ac{RE} (Figure \ref{fig:3}), was answered by 72 of 93 participants. The majority indicated that they either \textit{agree} or \textit{strongly agree} that they were interested in \ac{RE}. Results indicate that \sfrac{41}{72} (56.94\%) of responses \textit{strongly agree} and \sfrac{22}{72} (30.56\%) of responses \textit{somewhat agree}.
 
The mean percentage of time spent on \ac{RE} during work, shown in Appendix \hyperref[sec:appendixa]{A}, Table \hyperref[table:a2]{A.2} was found to be 26.56\% with a maximum of 90\%. The mean percentage of time spent \ac{RE} outside of work was found to be 27.02\% with a maximum of 80\%. Additionally when asked about how many years work has involved \ac{RE}, \sfrac{26}{83} (31.33\%) responses indicated that their career has not involved \ac{RE}, followed by \sfrac{13}{83} (15.66\%) responses indicating that their career has involved \ac{RE} for 5 to 6 years as shown in Appendix \hyperref[sec:appendixa]{A}, Table \hyperref[table:a3]{A.3}.

Participants then went into detail explaining how their work requires \ac{RE}. Many of the participants have worked on reversing malware either as part of an incidence response team or as \ac{RE} instructors. When explaining what \ac{RE} activities are done outside of work, most like to compete in \ac{CTF} competitions. Some are researches and study malware using \ac{RE}.

The professional roles participants have worked in are shown in Appendix \hyperref[sec:appendixa]{A}; Tables \hyperref[table:a4]{A.4}, \hyperref[table:a5]{A.5}. The responses indicated that  \sfrac{29}{344} (8.43\%) worked as engineers in the cybersecurity industry, \sfrac{27}{344} (7.85\%) worked as professors or educators while \sfrac{27}{344} (7.85\%) worked as software developers. It is important to note that the responses to this question varied indicating that participants held a diverse set of professional roles in the field of cybersecurity, and further exemplifying that \ac{RE} is a field that broadly applies to a number of job functions. It is important to note that the high volume of responses to this question (n=344), is due to participants selecting multiple options.

\subsection{User Education}
The proficiency for first \ac{RE} related roles (Appendix \hyperref[sec:appendixb]{B}; Table \hyperref[table:b1]{B.1}), shows \sfrac{28}{75} (37.33\%) responses indicating that they were not proficient, and only \sfrac{9}{75} (12.00\%) responses indicate they were very proficient followed by \sfrac{2}{75} (2.67\%) responses indicate they were an expert.

For where participants education in \ac{RE} took place (Appendix \hyperref[sec:appendixb]{B}; Table \hyperref[table:b2]{B.2}), \sfrac{57}{92} (61.96\%) indicated that they were self-taught, \sfrac{23}{92} (25.00\%) responses indicated they were taught from university or college, and \sfrac{12}{92} (13.04\%) responses indicate that they were taught by their employer. It is important to note that a significant amount of these participants indicated that they were not educated in \ac{RE} via a university, college, or employer.

Participants were then asked which resources were beneficial when learning \ac{RE} (Appendix \hyperref[sec:appendixb]{B}; Table \hyperref[table:b3]{B.3}). The majority of the responses indicated that they utilized  \textit{online instructional guides}, \textit{online challenges or ``crackme''}, and \textit{textbooks}. \sfrac{45}{213} (21.13\%) responses indicates that they utilized online instructional guides, \sfrac{40}{213} (18.78\%) responses indicates that they utilized online challenges or ``crackmes'', and \sfrac{36}{213} (16.90\%) responses indicated that they employed textbooks. It is important to note that the high volume of responses to this question (n=213), is due to some participants selecting multiple options. 

Most valuable books for learning \ac{RE} had (Appendix \hyperref[sec:appendixb]{B}; Table \hyperref[table:b5]{B.5}) majority of participants indicating that \textit{Practical Malware Analysis : The Hands-On Guide to Dissecting Malicious Software} and \textit{The \ac{IDA}/\ac{IDA} Pro Book} were the most valuable. \sfrac{34}{143} (23.78\%) responses indicated \textit{Practical Malware Analysis : The Hands-ON Guide to Dissecting Malicious Software} and \sfrac{23}{143} (16.08\%) responses indicated \textit{The \ac{IDA}/\ac{IDA} Pro Book}. It is important to note that the high volume of responses to this question (n=143), is due to some participants selecting multiple options.

Most valuable websites for learning \ac{RE} resulted with (Appendix \hyperref[sec:appendixb]{B}; Table \hyperref[table:b4]{B.4}) majority of participants indicating that \textit{youtube.com}, \textit{sans.com}, and \textit{crackmes.one} were most valuable. \sfrac{36}{114} (31.58\%) responses indicated \textit{youtube.com}, \sfrac{16}{114} (14.04\%) responses indicated \textit{sans.com}, and \sfrac{16}{114} (14.04\%) responses indicated \textit{crackmes.one}. It is important to note that the high volume of responses to this question (n=114), is due to participants selecting multiple options.

With the agreement if there is a shortage of effective \ac{RE} educational resources (Figure \ref{fig:3}), majority of participants indicated that they either \textit{strongly agree} or \textit{somewhat agree}. \sfrac{24}{71} (33.80\%) responses indicated strongly agree and \sfrac{22}{71} (30.99\%) responses indicated somewhat agree. Additionally, when asked if participants agree that cybersecurity graduates are leaving school with adequate proficiency in \ac{RE} (Figure \ref{fig:3}), majority of participants indicated that they \textit{somewhat disagree} or \textit{strongly disagree}. \sfrac{27}{72} (37.50\%) responses indicate somewhat disagree and \sfrac{17}{72} (23.61\%) responses indicated strongly disagree. It should also be noted that only \sfrac{4}{72} (5.56\%) responses indicated strongly agree. Furthermore, when asked if participants agree that there is a shortage of qualified \ac{RE} candidates in the workforce (Figure \ref{fig:3}), majority of participants indicated that they either \textit{strongly agree} or \textit{somewhat agree}. \sfrac{33}{71} (46.48\%) responses indicated strongly agree and \sfrac{20}{71} (28.17\%) responses indicated somewhat agree. It is also important to note that none of the participants indicated that they strongly disagree.

\subsection{Reverse Engineering Tools}
\begin{figure}[ht]
\centering
\begin{tikzpicture}[thick,font=\footnotesize]
\begin{axis}[
title=Knowledge and Usage of RE Tools,
    xbar,
    ylabel={Tools},
    xlabel={Percentage},
    y=4mm,
    bar width=1mm,
    symbolic y coords={Other,Qira,Apk2Gold,Objection,Boomerang,Krakatau,Uncompyle,Barf,Androguard,Triton,Jeb,PEDA,Jadx,CheatEngine,Angr,CFFExplorer,Frida,Hopper,Remnux,Dex2jar,Objdump,Binwalk,CuckooSandbox,BinaryNinja,Radare2,YARA,ApkTool,GDB,Ghidra,IDA/IDAPro},
    ytick=data, 
	width=7cm,height=15cm,
    legend style={at={(0.86,0.94)},anchor=south},
   yticklabel style = {font=\small},
   xticklabel={$\pgfmathprintnumber{\tick}\%$}
    ]

\addplot coordinates {(7.59,IDA/IDAPro) (7.46,Ghidra) (6.09,GDB) (5.85,ApkTool) (5.60,YARA) (5.47,Radare2) (5.35,BinaryNinja) (5.10,CuckooSandbox) (5.10,Binwalk) (5.10,Objdump) (4.35,Dex2jar) (4.23,Remnux) (3.73,Hopper) (3.61,Frida) (3.48,CFFExplorer) (2.86,Angr) (2.61,CheatEngine) (2.24,Jadx) (2.11,PEDA) (1.87,Jeb) (1.62,Triton) (1.62,Androguard) (1.24,Barf) (1.12,Uncompyle) (0.75,Krakatau) (0.75,Boomerang) (0.62,Objection) (0.62,Apk2Gold) (0.37,Qira) (1.49,Other)};

\addplot coordinates {(13.11,IDA/IDAPro) (8.74,Ghidra) (7.97,GDB) (3.86,ApkTool) (5.14,YARA) (5.14,Radare2) (3.60,BinaryNinja) (5.14,CuckooSandbox) (6.43,Binwalk) (7.46,Objdump) (5.14,Dex2jar) (4.37,Remnux)	(1.54,Hopper) (1.80,Frida) (4.63,CFFExplorer) (2.57,Angr) (1.80,CheatEngine) (2.31,Jadx) (2.83,PEDA) (0.51,Jeb)
(0.51,Triton) (0.51,Androguard) (0.51,Barf) (1.54,Uncompyle) (0.26,Krakatau) 		(0.51,Boomerang) (0.51,Objection) (0.00,Apk2Gold) (0.00,Qira) (1.54,Other)};

\legend{Knowledge,Usage}
\end{axis}
\end{tikzpicture}
\caption{participants knowledge in \ac{RE} tools compared to their usage of \ac{RE} tools.}
\label{fig:4}
\end{figure}
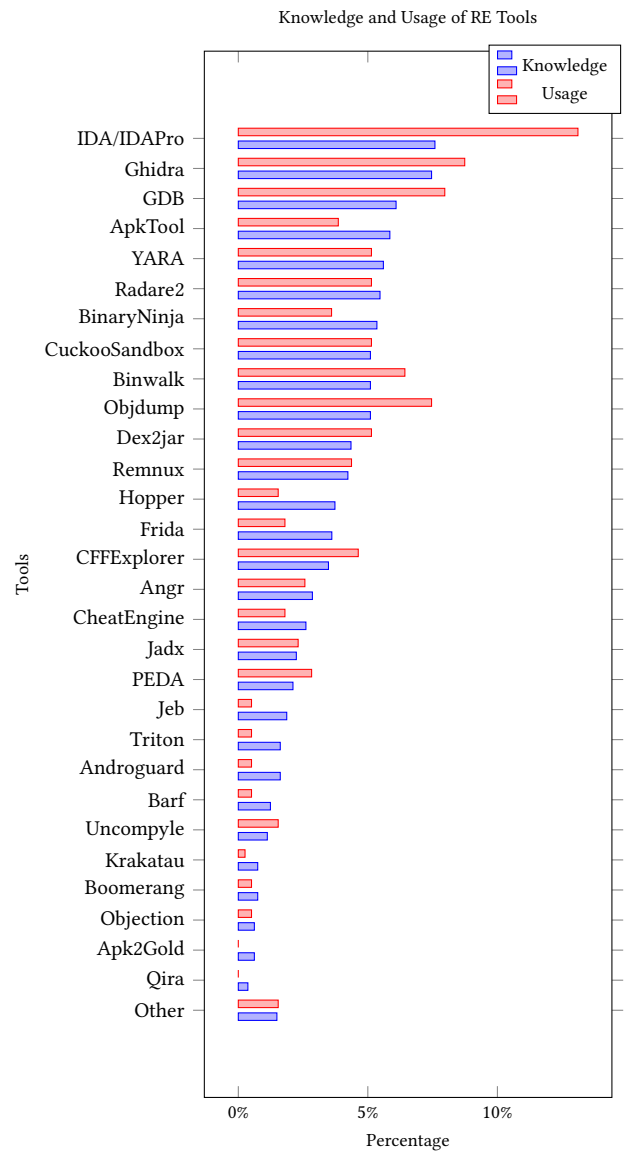

From the \ac{RE} tools participants had heard of (Figure \ref{fig:4}), majority heard of \ac{IDA}/\ac{IDA} Pro, Ghidra, and \ac{GDB}. \sfrac{61}{804} (7.59\%) responses indicated \ac{IDA}/\ac{IDA} Pro, \sfrac{60}{804} (7.46\%) responses indicated Ghidra, and \sfrac{49}{804} (6.09\%) responses indicating \ac{GDB}. It is important to note that per the high volume of responses to this question (n=804), some participants selected multiple options. When asked which tools participants use in \ac{RE} (Figure \ref{fig:4}), the majority indicated that they made use of \ac{IDA}/\ac{IDA} Pro, \ac{GDB}, and Ghidra. \sfrac{51}{389} (13.11\%) responses indicated \ac{IDA}/\ac{IDA} Pro, \sfrac{34}{389} (8.74\%) indicated Ghidra, and \sfrac{31}{389} (7.97\%) indicated \ac{GDB}. It is important to note that the high volume of responses to this question (n=389), is due to participants selecting multiple options.

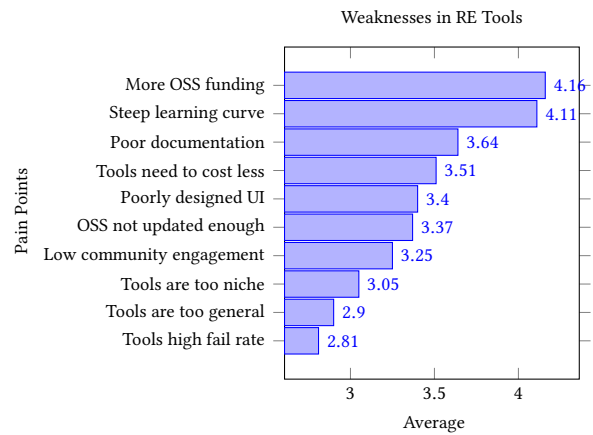
\begin{figure}[ht]
\centering
\begin{tikzpicture}[thick,font=\footnotesize]
\begin{axis}
[  title=Weaknesses in RE Tools,
    xbar,  
    enlargelimits=0.15,  
    xlabel={\ Average}, 
    ylabel={\ Pain Points},  
    width=5.5cm,
    height=6cm,
    symbolic y coords={Tools high fail rate, Tools are too general, Tools are too niche, Low  community  engagement, OSS not updated enough, Poorly designed UI, Tools need to cost less, Poor documentation, Steep learning curve, More OSS funding},
    ytick=data, 
     nodes near coords,
    nodes near coords align={horizontal}, 
    ]  
\addplot coordinates {(4.16,More OSS funding) (4.11,Steep  learning  curve) (3.64,Poor documentation) (3.51,Tools need to cost less) (3.40,Poorly designed UI) (3.37,OSS not updated enough) (3.25,Low community engagement) (3.05,Tools are too niche) (2.9,Tools are too general) (2.81,Tools high fail rate)};  
  \todo{Abe: Figure 2 does not look good at all}
\end{axis}  
\end{tikzpicture}
\caption{Pain points in regards to \ac{RE} tools. Note:  The average is calculated on a scale from 1 - 5. Strongly Disagree = 1 and Strongly Agree = 5}
\label{fig:1}
\end{figure}

Participants were then asked if the tools that they use for \ac{RE} are adequate for the work they need to accomplish (Figure \ref{fig:3}). The majority either \textit{somewhat agree} or \textit{strongly agree}. \sfrac{27}{65} (41.54\%) responses \textit{somewhat agree} and \sfrac{21}{65} (32.31\%) responses \textit{strongly agree}. When asked what some weaknesses in regards to current \ac{RE} tools are (Figure~\ref{fig:1}), based upon the mean responses most participants agreed that the primary pain points are the steep learning curve to use tools, open-source software (OSS) need more funding, tools need to cost less, and poor documentation of tools.

The operating systems most utilized by the participants for \ac{RE} (Appendix \hyperref[sec:appendixc]{C}; Table \hyperref[table:c1]{C.1}), were either Unix/Linux and Windows. \sfrac{52}{124} (41.94\%) responses indicated utilize Unix/Linux and \sfrac{47}{124} (37.90\%) indicated Windows. It is important to note that the high volume of responses to this question (n=124), is due to participants selecting multiple options. 

The majority of architectures encountered in \ac{RE} (Appendix \hyperref[sec:appendixc]{C}; Table \hyperref[table:c2]{C.2}), were x86/i386, x86-64/AMD64, and \ac{ARM}. \sfrac{50}{131} (38.17\%) responses indicated x86/AMD64, \sfrac{45}{131} (34.35\%) indicated x86/i386 , and \sfrac{24}{131} (18.32\%) indicated \ac{ARM}. It is important to note that the high volume of responses to this question (n=131), is due to participants selecting multiple options. When asked which platforms participants have encountered while \ac{RE} (Appendix \hyperref[sec:appendixc]{C}; Table \hyperref[table:c3]{C.3}), the majority indicated that they had encountered Windows, *nix, and Android. \sfrac{53}{131} (40.46\%) responses indicated Windows, \sfrac{39}{131} (29.77\%) indicated *nix, and \sfrac{29}{131} (22.14\%) indicated Android. It is important to note that the high volume of responses to this question (n=131), is due to participants selecting multiple options.

Languages that participants have encountered while \ac{RE} (Appendix \hyperref[sec:appendixc]{C}; Table \hyperref[table:c4]{C.4}), indicated that majority had encountered C/C++, Java, Compiled Python, and C\#. \sfrac{58}{193} (30.05\%) responses indicated C/C++, \sfrac{39}{193} (20.21\%) indicated Java, \sfrac{33}{193} (17.10\%) indicated C\# and \sfrac{30}{193} (15.54\%) indicated Compiled Python. It is important to note that the high volume of responses to this question (n=193), is due to participants selecting multiple options. When asked which software/systems participants have reverse engineered or encountered whilst \ac{RE} (Appendix \hyperref[sec:appendixc]{C}; Table \hyperref[table:c5]{C.5}), the majority indicated malware and desktop applications. \sfrac{50}{176} (28.41\%) responses indicated malware and \sfrac{40}{176} (22.73\%) indicated desktop applications. It is important to note that the high volume of responses to this question (n=176), is due to participants selecting multiple options. 

Participants elaborated on their experience on learning \ac{RE}. Many learned though academic courses, \ac{CTF} challenges, online tutorials and a lot of practice on their own. Reading was said to only help understand general concepts. Anything other than hands on experience people don't find helpful. Not everyone started with the assembly languages when learning \ac{RE}. There were people who started by understanding Python and how C compiles to assembly. Many found that it was much harder if you didn't have experience in programming before starting to learn \ac{RE}.
\subsection{Ghidra vs IDA}

\begin{figure}[ht]
\centering
\begin{tikzpicture}[thick,font=\footnotesize]
\begin{axis}[
title=Effectiveness of Features Between Tools,
    xbar,
    enlargelimits=0.15,
    xlabel={Average},
    ylabel={Features},
    bar width=1.5mm,
    width=5cm,
    height=7cm,
    symbolic y coords={Third-party plugins, Debugger, Library func. sig. recog, Hex view, Function signature mod., Reference table, Function call graph, Variable renaming, Graph view, Disassembler},
    legend style={at={(-0.5,0.05)},
    anchor=north,legend columns=-1},
    ytick=data, 
    nodes near coords,
    nodes near coords align={horizontal}
    ]
\addplot coordinates {(4.39,Disassembler) (4.22,Graph view) (3.94,Variable renaming) (3.94,Function call graph) (3.88,Reference table) (3.85,Function signature mod.) (3.69,Hex view) (3.67,Library func. sig. recog) (3.56,Debugger) (3.55,Third-party plugins)};

\addplot coordinates {(4.41,Disassembler) (4.33,Graph view) (4.31,Variable renaming) (4.06,Function call graph) (4.00,Reference table) (3.94,Function signature mod.)
(3.88,Hex view) (3.81,Library func. sig. recog) (3.69,Debugger) (3.63,Third-party plugins)};

\legend{Ghidra,IDA}

\end{axis}
\end{tikzpicture}
\caption{Effectiveness of features between \ac{IDA}/\ac{IDA} Pro and Ghidra for \ac{RE} tasks. Note:  The average is calculated on a scale from 1 - 5. Strongly Disagree = 1 and Strongly Agree = 5}
\label{fig:2}
\end{figure}
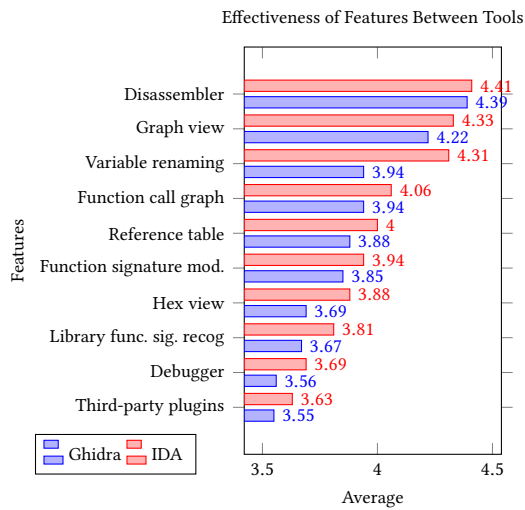

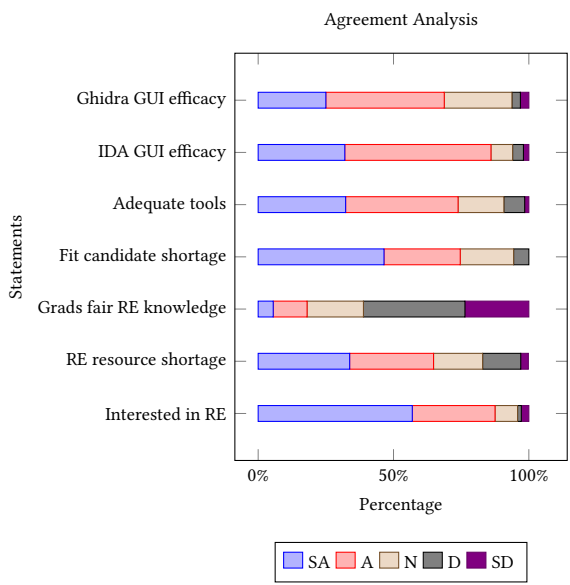
\begin{figure}[ht]
\centering
\begin{tikzpicture}[thick,font=\footnotesize]
\begin{axis}[
title=Agreement Analysis,
    xbar stacked,
	bar width=6pt,
	enlargelimits=0.15,
    width=6cm,
    height=7cm,
    legend style={at={(0.5,-0.20)},
      anchor=north,legend columns=-1},
    xlabel={Percentage},
    ylabel={Statements},
    symbolic y coords={Interested in RE, RE resource shortage, Grads fair RE knowledge, Fit candidate shortage, 
		Adequate tools, IDA GUI efficacy, Ghidra GUI efficacy},
    ytick=data,
     xticklabel={$\pgfmathprintnumber{\tick}\%$}
    ]
\addplot coordinates {(56.94,Interested in RE) (33.80,RE resource shortage) (5.56,Grads fair RE knowledge) (46.48,Fit candidate shortage) (32.31,Adequate tools) (32,IDA GUI efficacy) (25,Ghidra GUI efficacy)};
\addplot coordinates {(30.56,Interested in RE) (30.99,RE resource shortage) (12.50,Grads fair RE knowledge) (28.17,Fit candidate shortage) (41.54,Adequate tools) (54,IDA GUI efficacy) (43.75,Ghidra GUI efficacy)};
\addplot coordinates {(8.33,Interested in RE) (18.13,RE resource shortage) (20.83,Grads fair RE knowledge) (19.72,Fit candidate shortage) (16.92,Adequate tools) (8,IDA GUI efficacy) (25,Ghidra GUI efficacy)};
\addplot coordinates {(1.39,Interested in RE) (14.08,RE resource shortage) (37.50,Grads fair RE knowledge) (5.63,Fit candidate shortage) (7.69,Adequate tools) (4,IDA GUI efficacy) (3.13,Ghidra GUI efficacy)};
\addplot coordinates {(2.78,Interested in RE) (2.82,RE resource shortage) (23.61,Grads fair RE knowledge) (0,Fit candidate shortage) (1.54,Adequate tools) (2,IDA GUI efficacy) (3.13,Ghidra GUI efficacy)};

\legend{\strut SA, \strut A, \strut N, \strut D, \strut SD}
\end{axis}
\end{tikzpicture}
\caption{Results from agreement questions. Note: Strongly Agree (SA), Somewhat Agree (A), Neither (N), Somewhat Disagree (D), Strongly Disagree (SD)}
\label{fig:3}
\end{figure}

Whether participants use the free or commercial edition of \ac{IDA} (Appendix \hyperref[sec:appendixd]{D}; Table \hyperref[table:d1]{D.1}), there was an even split of 25 responses for both. The majority of responses agreed that disassembler, variable renaming, graph view, function call graph, and reference table were the most effective features of \ac{IDA}/\ac{IDA} Pro for \ac{RE} (Figure~\ref{fig:2}). When asked which \ac{IDA}/\ac{IDA} Pro \ac{GUI} features participants believe could be improved (Appendix \hyperref[sec:appendixd]{D}; Table \hyperref[table:d2]{D.2}), majority of responses indicated debugger, third-party plugins, and disassembler. \sfrac{18}{60} (30\%) responses indicated debugger, \sfrac{11}{60} (18.33\%) indicated third-party plugins, and \sfrac{9}{60} (15\%) indicated disassembler. When asked if participants agree that \ac{IDA}/\ac{IDA} Pro's \ac{GUI} is effective for \ac{RE} (Figure \ref{fig:3}), majority of \sfrac{27}{50} (54\%) responses indicated \textit{somewhat agree}. Lastly, when asked if participants believe that \ac{IDA}/\ac{IDA} Pro has any features which are advantageous over those in Ghidra (Appendix \hyperref[sec:appendixd]{D}; Table \hyperref[table:d3]{D.3}), majority of \sfrac{14}{27} (51.85\%) responses indicated no.

For features of Ghidra that are effective for \ac{RE} (Figure~\ref{fig:2}), the majority agreed on: disassembler, variable renaming, reference table, function signature modification, and data type viewer. When asked which features of Ghidra could be improved (Appendix \hyperref[sec:appendixd]{D}; Table \hyperref[table:d4]{D.4}), majority of responses indicated debugger and graph view. \sfrac{9}{43} (20.93\%) responses indicated debugger and \sfrac{8}{43} (18.60\%) indicated graph view. When asked if participants agree that Ghidra's \ac{GUI} is effective in representing information (Figure \ref{fig:3}), majority of \sfrac{14}{32} (43.75\&) responses indicated somewhat agree. It is important to note that participants could select multiple options.

Lastly, when asked if participants believe that Ghidra has any features which are better than those in \ac{IDA}/\ac{IDA} Pro (Appendix \hyperref[sec:appendixd]{D}; Table \hyperref[table:d5]{D.5}), \sfrac{13}{25} (52\%) of responses indicated yes and \sfrac{12}{25} (48\%) of responses indicated no.

\subsection{Teaching Reverse Engineering}
The question if participants teach or have ever taught \ac{RE} in a classroom setting (Appendix \hyperref[sec:appendixe]{E}; Table \hyperref[table:e1]{E.1}), majority of \sfrac{48}{72} (66.67\%) responses indicated that they have not, whereas \sfrac{24}{72} (33.33\%) indicated that they have. 
 
The resources participants used to teach \ac{RE} (Appendix \hyperref[sec:appendixe]{E}; Table \hyperref[table:e2]{E.2}), had a majority indicating that they utilized demonstrations and lab activities. \sfrac{20}{72} (27.28\%) responses indicated that they utilized lab activities and \sfrac{19}{72} (26.39\%) indicated they utilized demonstrations. Additionally, when asked what types of assignments they gave to their students to teach \ac{RE} (Appendix \hyperref[sec:appendixe]{E}; Table \hyperref[table:e3]{E.3}), majority of responses indicated that they utilized projects and \ac{CTF} style assignments. \sfrac{15}{40} (37.50\%) responses indicated that they utilized projects and \sfrac{15}{40} (37.50\%) indicated that they utilized \ac{CTF} style assignments. Furthermore, when asked what methods participants use to assess their students' knowledge of \ac{RE} (Appendix \hyperref[sec:appendixe]{E}; Table \hyperref[table:e4]{E.4}), majority of responses indicated in-class presentations, exams and projects/labs. \sfrac{8}{32} (25.00\%) responses indicated that they utilize in-class presentations, \sfrac{7}{32} (21.88\%) indicated that they utilize exams, and \sfrac{7}{32} (21.88\%) indicated that they utilize other methods, most of which were either labs or projects. It is important to note that participants could select multiple options. 

In the agreement that certain circumstances are challenges in teaching \ac{RE} (Appendix \hyperref[sec:appendixe]{E}; Table \hyperref[table:e5]{E.5}), majority of participants agreed on three circumstances. \sfrac{18}{61} (29.51\%) responses agreed on students' lack of prerequisite knowledge, \sfrac{13}{61} (21.31\%) agreed on students' understanding of assembly language, and \sfrac{11}{61} (18.03\%) agreed on the learning curve of \ac{RE}.

Lastly, when asked what tools are used in classes to teach \ac{RE} (Appendix \hyperref[sec:appendixe]{E}; Table \hyperref[table:e6]{E.6}), majority of participants used \textit{\ac{IDA}/\ac{IDA} Pro}, \textit{\ac{GDB}}, \textit{Ghidra} and \textit{objdump}. \sfrac{14}{98} (14.29\%) responses indicated \textit{\ac{IDA}/\ac{IDA} Pro}, \sfrac{10}{98} (10.20\%) indicated \textit{\ac{GDB}}, \sfrac{10}{98} (10.2\%) indicated \textit{Ghidra}, and \sfrac{10}{98} (10.2\%) indicated \textit{Objdump}. It is important to note that the high volume of responses to this question (n=98), is due to participants selecting multiple options.

\subsection{Web-Based Reverse Engineering}
Participants were asked if they knew of an existing web-based \ac{RE} platform (Appendix \hyperref[sec:appendixf]{F}; Table \hyperref[table:f1]{F.1})and the majority of \sfrac{49}{63} (77.78\%) responses indicated no, but \sfrac{14}{63} (22.22\%) indicated yes. The web-based \ac{RE} platforms participants knew of include the following: Acunetix, Binary Ninja Cloud, \ac{IDA},  and radare2.

The interest of a web-based \ac{RE} platform (Appendix \hyperref[sec:appendixf]{F}; Table \hyperref[table:f2]{F.2}), were the majority of maybe and yes. \sfrac{31}{65} (47.69\%) responses indicated maybe and \sfrac{24}{65} (36.92\%) indicated yes. Lastly, when asked what characteristics or aspects of such a web-based \ac{RE} platform participants would like to see (Appendix \hyperref[sec:appendixf]{F}; Table \hyperref[table:f3]{F.3}), majority of responses indicated access to various tools/features and allows for collaboration with other users. \sfrac{46}{216} (21.30\%) of responses indicated access to various tools/features and \sfrac{40}{216} (18.52\%) responses indicated allowing for collaboration with other users. It is important to note that the high volume of responses to this question (n=216), is due to participants selecting multiple options.

\subsection{Challenges}
Whether there are legal challenges when it comes to \ac{RE} the majority said no. Some participants mention that they get permission from the authors of the software before reversing it. Participants also mentioned how the \ac{DMCA} and \ac{EULA} limits what can be used as an \ac{RE} task.

With future challenges , the majority see malware and software becoming more complex. In particular, encryption and other software protections will be getting more advanced. Some see this rapid growth as a negative as their are already a lack of experts. It will also be harder for beginners who don't have much practical knowledge.

%% file: sections/sec_5.tex
\section{Limitations} \label{sec:limitations}
Our target audience included students, professionals, and educators that have \ac{RE} experience. When trying to find participants, it was found that this area of cybersecurity isn't as popular and the understanding is still relatively new. This resulted in fewer number of participants, even though overall, we received an acceptable sample size. Throughout the survey each question was optional allowing participants to skip and leave questions blank. It became clear that some questions were favored over others which resulted in inconsistent response counts for each question. There were also three participants that dropped out after the definitions portion and seven participates that stopped after the first question. This could be the result of loss of interest from the length of the survey with fifty-eight questions.

%% file: sections/sec_6.tex
\section{Related Work}
\label{sec:related work}
Extensive work has been conducted on how to best engage computer science students and on the best methods for  teaching cybersecurity concepts and methodologies. Work has been done in identifying the learning styles of computer science and engineering students in order to determine the best teaching methods \citep{felder1988learning, thomas2002learning}. 

\ac{CTF} style challenges are created for the learning benefit of students. They consist of categories like web exploitation, cryptography and binary \ac{RE}. The creators of \ac{CTF} competitions create these to teach old and new techniques. The website \textit{https://tryhackme.com/} has many modules we're you can learn and practice \ac{RE} concepts as well as other cybersecurity subjects in a \ac{CTF} like fashion. Other works have tried to incorporate \ac{CTF} style labs into education for methods of teaching cybersecurity principals \citep{cheung2011challenge, chothia2015offline, olano2014securityempire}, which has also been used to teach \ac{RE} concepts \citep{feng2015scaffolded}.

A number of educators have documented their approach to creating and teaching \ac{RE} courses \citep{barr2000introduction, richard2009highly}. 

Despite the fact that 21 higher education institutions \citep{caeco} have been designated as \ac{CAE-CO} institutions by the \ac{NSA}, which requires a course in \ac{RE}, there is still a very large knowledge gap in individuals that can effectively reverse engineer.

Another way in which students have approached learning cybersecurity concepts is through self-taught methods such as certifications. The problem with this is that there is a severe lack of resources for individuals who decide to take this approach.  Given the prerequisite knowledge necessary for \ac{RE}, it is very difficult for beginners to get started. While there are a number of avenues available to individuals interested in learning, e.g., programming, such as online interactive teaching tools, among the few ``beginner'' resources in \ac{RE} that are available, most are very intimidating \citep{ReverseEngineeringForBeginners}.

\section{Discussion and Conclusions}
\label{sec:discussion/conclusions}

The results from our work illustrates that this domain is at its infancy with few experts and a lack of proficiency. The results also illustrate a lack of diversity amongst \ac{RE} professionals. Among the participants that had an understanding of \ac{RE}, a focus in cybersecurity or \ac{RE}, could explain why most participants have either a masters or doctorate degree. Further development of courses at a basic level for undergraduate students and activities for high school students could prove to be beneficial. An early introduction to \ac{RE} to students during middle school and high school may open up new interests that could turn into future academic focuses. The data suggests that computer science is the most popular field of study, perhaps due to it being the oldest major, or due to the advanced knowledge of programming needed when performing \ac{RE}. Being self-taught turned out to be how most learned \ac{RE}. This could also be why many respondents enjoy \ac{RE} as a hobby just as much when working at their job.   

Furthermore, another important conclusion made from the results of this survey is that the number of tools individuals use in \ac{RE} are limited: the two most popular being \ac{IDA}/\ac{IDA} Pro and Ghidra, which were followed by \ac{GDB} and Objdump as the next most popular. This popularity could be the result of the age of these tools, although Ghidra was only made public recently, which may hint that the maturity and ease of use of the tools may be a more important factor in their adoption in \ac{RE}. One interesting item to note is that more participant claimed to use \ac{IDA}/\ac{IDA} Pro, than actually considering themselves knowledgeable in \ac{IDA}/\ac{IDA} Pro. In fact the trend as seen in Figure \ref{fig:4} is that the popular tools for \ac{RE} are generally used more than they are understood which may indicate an underlying issue in \ac{RE} educational approaches. This could be the result of the lack of schools with \ac{RE} courses, resulting in the only option being to teach yourself and find other resources through either books or online resources. It may be helpful to add lectures or a playlist of video tutorials to already existing \ac{RE} courses on the usage and functionality of the more popular tools. 

Additionally, on the front of \ac{IDA} versus Ghidra, it is important to note, as shown in Figure \ref{fig:2}, that \ac{IDA} ranks higher than Ghidra across the board, in some features more than others. The feature preferred in \ac{IDA} over Ghidra the most was variable renaming, this was the only feature that did not have  minimal differences in the preference between the two \ac{RE} tools which makes it clear that there should probably be work done to improve variable renaming in Ghidra. We also found it interesting that exactly half of respondents indicated their use of the free edition of \ac{IDA} where as the other half used the commercial edition. This may represent an even split in the demand for \ac{RE} tools in a commercial setting v.s. everyday, or minimal functionality usage. Although it should also be noted that when asked about if \ac{IDA}/\ac{IDA} Pro was advantageous over Ghidra based upon their feature sets, a slim majority indicated no.

The results showed that most who have taught \ac{RE} prefer to use hands on activities and labs. This is a step in the positive direction as it tells us that instructors may have found an optimal teaching approach. The lack of a student's prerequisite knowledge may be a result of \ac{RE} classes being an elective, rather than a requirement. \ac{RE} is a more advanced class that requires knowledge in multiple areas. The steep learning curve may discourage new students away, which could explain the lack of experts found in the results. \ac{RE} is a field in the workforce that is very much needed for incident response teams.

The researchers acknowledge that a follow-up survey or additions to the current survey would be beneficial to gain more insight on specifics. Those that indicated they disagreed with certain question statements, could have also been asked to provide an explanation. With all surveys, there needs to be an understanding of what is wrong and what solutions must be made. There were also questions that participants could select tools to indicate their usage or efficacy. Future work could allow participants to rank the chosen tools. Lastly, a longitudinal study that includes both surveys and interviews with experts may shed light on how the domain and its educational capacity is changing over time. This is imperative as \ac{RE} is becoming a cornerstone for improving the cybersecurity and incidence response stature of public and private organizations.

%% file: sections/appendix_a.tex
\clearpage
\makeatletter
\setlength{\@fptop}{33pt}
\makeatother



\twocolumn[\section*{\hfil Appendix A. Demographics\hfil}]
\label{sec:appendixa}

\NetTableWidth=\dimexpr
    \linewidth
    - 8\tabcolsep
    - 5\arrayrulewidth
\relax

\begin{table}[!h]
\centering
\caption*{Table A.1. Demographics.}
\begin{tabular}{p{.64\NetTableWidth} p{.12\NetTableWidth} p{.24\NetTableWidth}} 
\hline
 & \textbf{Count} & \textbf{Percentage} \\ \hline
\textbf{Race} \\
White & 56 & 65.88\% \\
Do not wish to specify & 13 & 15.29\% \\
Asian & 11 & 12.94\% \\
Black or African American & 1 & 1.18\% \\
Other & 4 & 4.71\% \\
- Israeli & & \\
- White, African American, and American Indian & & \\
\hline
\textbf{Sex} \\
Male & 69 & 81.18\% \\
Female & 10 & 11.76\% \\
Do not wish to specify & 6 & 7.06\% \\
\hline
\textbf{Age} \\
30-39 & 30 & 35.29\% \\
20-29 & 24 & 28.24\% \\
40-49 & 12 & 14.12\% \\
Younger than 20 & 6 & 7.06\% \\
50-59 & 5 & 5.88\% \\
60-69 & 4 & 4.71\% \\
Do not wish to specify & 3 & 3.53\% \\
Older than 60 & 1 & 1.18\% \\
\hline
\textbf{Level of Education} \\
Master's degree & 24 & 28.24\% \\
Doctoral degree & 18 & 21.18\% \\
Bachelor's degree & 17 & 20.00\% \\
Some university or college & 10 & 11.76\% \\
High school diploma or GED & 6 & 7.06\% \\
Associate's degree & 5 & 5.88\% \\
Some high school & 5 & 5.88\% \\
\hline
\textbf{Field of Study} \\
Computer Science & 26 & 31.33\% \\
Cybersecurity & 24 & 28.92\% \\
Computer Engineering & 9 & 10.84\% \\
Information Technology & 7 & 8.43\% \\
Other & 5 & 6.02\% \\
Electrical Engineering & 3 & 3.61\% \\
Criminal Justice & 2 & 2.41\% \\
Mechanical Engineering & 1 & 1.20\%	\\
Mathematics & 1 & 1.20\% \\
Computer Networks & 1 & 1.20\% \\
Civil Engineering & 1 & 1.20\% \\
Business Administration & 1 & 1.20\% \\
Biology & 1 & 1.20\% \\
\hline
\end{tabular}
\vspace{0.1in}
\label{table:a1}

\centering
\caption*{Table A.2. Percentage of time related to reverse engineering.}
\begin{tabular}{p{.5\NetTableWidth} p{.5\NetTableWidth}}
\hline
\textbf{Location} & \textbf{Average Percentage} \\ \hline
During work & 26.56\% \\
Outside work & 27.02\% \\\hline
\end{tabular}
\label{table:a2}

\end{table}

\begin{table}[!ht]
\centering
\caption*{Table A.3. Years of reverse engineering during career.}
\begin{tabular}{p{.70\NetTableWidth} p{.15\NetTableWidth} p{.25\NetTableWidth}}
\hline
\textbf{Years} & \textbf{Count} & \textbf{Percentage} \\ \hline
My career has not involved reverse engineering & 26 & 31.33\%\\
5-6 & 13 & 15.66\%\\
3-4 & 12 & 14.46\%\\
Less than 1 & 10 & 12.05\%\\
1-2 & 10 & 12.05\%\\
More than 10 & 6 & 7.23\%\\
7-8 & 4 & 4.82\%\\
9-10 & 2 & 2.41\%\\\hline
\end{tabular}
\label{table:a3}

\begin{center}
\caption*{Table A.4. Professional Roles.}
\begin{tabular}{p{.75\NetTableWidth} p{.12\NetTableWidth} p{.24\NetTableWidth}}
\hline
\textbf{Role} & \textbf{Count} & \textbf{Percentage} \\ \hline
Engineer in Cybersecurity Industry & 29 & 8.43\% \\
Professor or Educator & 27 & 7.85\% \\
Software Developer & 27 & 7.85\% \\
Forensic Analyst & 22 & 6.40\% \\
System Administrator & 19 & 5.52\% \\
Cyber Instructor & 16 & 4.65\% \\
Cyber Instructional Curriculum Developer & 12 & 3.49\% \\
Director or Management & 11 & 3.20\% \\
Research \& Development Specialist & 11 & 3.20\% \\
Cyber Defense Analyst & 11 & 3.20\% \\
Cyber Defense Incident Responder & 11 & 3.20\% \\
Vulnerability Analyst & 11 & 3.20\% \\
Cyber Defense Forensics Analyst & 11 & 3.20\% \\
Security Architect & 9 & 2.62\% \\
Testing and Evaluation Specialist & 9 & 2.62\% \\
Systems Developer & 9 & 2.62\% \\
Cyber Crime Investigator & 9 & 2.62\% \\
Network Operations Specialist & 8 & 2.33\% \\
Database Administrator & 7 & 2.03\% \\
System Security Analyst & 7 & 2.03\% \\
Information Systems Security Developer & 6 & 1.74\% \\
Technical Support Specialist & 6 & 1.74\% \\
Secure Software Assessor & 4 & 1.16\% \\
IT Project Manager & 4 & 1.16\% \\
Exploitation Analyst & 4 & 1.16\% \\
\hline
\end{tabular}
\end{center}
\label{table:a4}
\end{table}

\makeatletter
\setlength{\@fptop}{0pt}
\makeatother

\begin{table*}[t!]
\begin{center}
\caption*{Table A.4. Professional Roles (Continued).}
\begin{tabular}{p{.75\NetTableWidth} p{.12\NetTableWidth} p{.24\NetTableWidth}}
\textbf{Role} & \textbf{Count} & \textbf{Percentage} \\ \hline
Cyber Intel Planner & 4 & 1.16\% \\
Cyber Operator & 4 & 1.16\% \\
Executive & 3 & 0.87\% \\
Enterprise Architect & 3 & 0.87\% \\
Warnings Analyst & 3 & 0.87\% \\ 
All-Source Analyst & 3 & 0.87\% \\
Cyber Workforce Developer and Manager & 2 & 0.58\% \\
Program Manager & 2 & 0.58\% \\
Cyber Defense Infrastructure Support Specialist & 2 & 0.58\% \\
Security Control Assessor & 1 & 0.29\% \\
Requirements Planner & 1 & 0.29\% \\
Knowledge Manager & 1 & 0.29\% \\
Information Systems Security Manager & 1 & 0.29\% \\
Mission Assessment Specialist & 1 & 0.29\% \\
Target Developer & 1 & 0.29\% \\
Target Analyst & 1 & 0.29\% \\
Language Analyst & 1 & 0.29\% \\
Language Analyst & 1 & 0.29\% \\
All Source-Collection Manager & 1 & 0.29\% \\
All Source-Collection Requirements Evaluation Manager & 1 & 0.29\% \\
Cyber Operations Planner & 1 & 0.29\% \\
Other & 5 & 6.02\% \\
- Outdoor education & &\\
- Telecommunications & &\\
\hline
\end{tabular}
\end{center}
\label{table:a5}
\end{table*}

%% file: sections/appendix_b.tex
\twocolumn[\section*{\hfil Appendix B. User Education\hfil}]
\label{sec:appendixb}

\begin{table}[!h]
\begin{center}
\caption*{Table B.1. Proficiency in reverse engineering .}
\begin{tabular}{p{.64\NetTableWidth} p{.15\NetTableWidth} p{.25\NetTableWidth}}
\hline
\textbf{Proficiency} & \textbf{Count} & \textbf{Percentage} \\ \hline
Not Proficient & 28  & 37.33\% \\
Little Proficiency & 16  & 21.33\% \\
Somewhat Proficient  & 20  & 26.67\% \\
Very Proficient & 9  & 12.00\% \\
Expert & 2  & 2.67\% \\
\end{tabular}
\end{center}
\label{table:b1}
\vspace{-0.1in}
\begin{center}
\caption*{Table B.2. Source of reverse engineering education.}
\begin{tabular}{p{.64\NetTableWidth} p{.12\NetTableWidth} p{.24\NetTableWidth}}
\hline
\textbf{Education Source} & \textbf{Count} & \textbf{Percentage} \\ \hline
Self-taught & 57 & 61.96\% \\
University of College & 23 & 25.00\% \\
Employer & 12 & 13.04\% \\\hline
\end{tabular}
\end{center}
\label{table:b2}
\vspace{-0.1in}
\begin{center}
\caption*{Table B.3. Beneficial resources for learning reverse engineering.}
\begin{tabular}{p{.64\NetTableWidth} p{.12\NetTableWidth} p{.24\NetTableWidth}}
\hline
\textbf{Resource} & \textbf{Count} & \textbf{Percentage} \\ \hline
Online instructional guides & 45 & 21.13\% \\
Online challenges or "crackmes" & 40 & 18.78\% \\
Textbooks & 36 & 16.90\% \\
Online video tutorials  & 33 & 15.40\% \\
College or University lectures & 25 & 11.74\% \\
On-the-job training & 22 & 10.33\% \\
Employer-sponsored training & 12 & 5.63\% \\
\end{tabular}
\end{center}
\label{table:b3}
\end{table}

\begin{table}
\begin{center}
\caption*{Table B.4. Most valuable websites for learning reverse engineering.}
\begin{tabular}{p{.64\NetTableWidth} p{.12\NetTableWidth} p{.24\NetTableWidth}}
\hline
\textbf{Website} & \textbf{Count} & \textbf{Percentage} \\ \hline
youtube.com & 36 & 31.58\% \\
crackmes.one & 16 & 14.04\% \\
sans.com & 16 & 14.04\% \\
codeacademy.com & 8 & 7.02\% \\
begin.re & 6 & 5.26\% \\
compilerexplorer.com & 3 & 2.63\% \\
udemy.com & 3 & 2.63\% \\
datacamp.com & 3 & 2.63\% \\
leetcode.com & 1 & 0.88\% \\
hackerrank.com & 1 & 0.88\% \\
khanacademy.com & 0 & 0.00\% \\
Other & 9 & 14.52\% \\
- pwnable.kr &  &\\
- pluralsight.com &  & \\
- hackthebox.eu &  & \\
- tuts4you.com &  & \\
- rtn-team.cc &  & \\ 
- sharplab.io &  & \\ 
- google.com &  & \\
- stackoverflow.com &  &\\
- tuts4you.com &  & \\
- sans.org &  & \\
- overthewire.org & & \\
\end{tabular}
\end{center}
\label{table:b4}
\end{table}

\clearpage
\makeatletter
\setlength{\@fptop}{33pt}
\makeatother

\begin{table*}[!htb]
\vspace{-7in}
\small
\begin{center}
\caption*{Table B.5. Most valuable books for learning reverse engineering.}
\begin{tabular}{c c c}
\hline
\textbf{Book} & \textbf{Count} & \textbf{Percentage} \\ \hline
Practical Malware Analysis: The Hands-on Guide to Dissecting Malicious Software & 34 & 23.78\% \\
The IDA/IDA Pro Book & 23 & 16.08\% \\
The Art of Memory Forensics: Detecting Malware and Threats in Windows, Linux, and Mac Memory & 18 & 12.59\% \\
Hacking: The Art of Exploitation & 15 & 10.49\% \\
Reversing: Secrets of Reverse Engineering & 13 & 9.09\% \\
Practical Binary Analysis & 13 & 9.09\% \\
Modern X86 Assembly Language Programming: 32-bit, 64-bit, SSE, and AVX & 8 & 5.59\% \\
Mastering Reverse Engineering & 6 & 4.20\% \\
Reverse Engineering: Mechanisms, Structures, Systems \& Materials & 3 & 2.10\% \\
Mastering Reverse Engineering: Re-engineer Your Ethical Hacking Skills & 3 & 2.10\% \\
Other & 7 & 4.90\% \\
Windows Internals & & \\\hline

\end{tabular}
\end{center}

\label{table:b5}
\end{table*}

%% file: sections/appendix_c.tex
\clearpage

\twocolumn[\section*{\hfil Appendix C. Reverse Engineering Tools\hfil}]
\label{sec:appendixc}

\begin{table}[!ht]

\begin{center}
\caption*{Table C.1. Operating systems utilized in reverse engineering environment.}
\begin{tabular}{p{.64\NetTableWidth} p{.12\NetTableWidth} p{.24\NetTableWidth}}

\hline
 \textbf{Operating System} & \textbf{Count} & \textbf{Percentage} \\ \hline
Unix/Linux & 52 & 41.94\% \\
Windows  & 47 & 37.90\% \\
macOS & 21 & 16.94\% \\
Other & 4 & 3.25\% \\ \hline
\end{tabular}

\end{center}
\label{table:c1}

\begin{center}
\caption*{Table C.2. Architectures encountered when reverse engineering.}
\begin{tabular}{p{.64\NetTableWidth} p{.12\NetTableWidth} p{.24\NetTableWidth}}
\hline
 \textbf{Architecture} & \textbf{Count} & \textbf{Percentage} \\ \hline
x86-64/AMD64 & 50 & 38.17\% \\
x86/i386 & 45 & 34.35\% \\     
ARM & 24 & 18.32\% \\
MIPS & 10 & 7.63\% \\
RISC-V & 2 & 1.53\% \\\hline

\label{table:c2}

\end{tabular}
\end{center}

\begin{center}
\caption*{Table C.3. Platforms encountered when reverse engineering.}
\begin{tabular}{p{.64\NetTableWidth} p{.12\NetTableWidth} p{.24\NetTableWidth}}
\hline
 \textbf{Platform} & \textbf{Count} & \textbf{Percentage} \\ \hline
Windows	  & 53 & 40.46\% \\
*nix	  & 39 & 29.77\% \\
Android	  & 29 & 22.14\% \\
Apple iOS & 10 & 4.63\% \\\hline
\end{tabular}
\end{center}
\label{table:c3}

\begin{center}
\caption*{Table C.4. Languages encountered when reverse engineering.}
\begin{tabular}{p{.64\NetTableWidth} p{.12\NetTableWidth} p{.24\NetTableWidth}}
\hline
 \textbf{Language} & \textbf{Count} & \textbf{Percentage} \\ \hline
C/C++       	& 58&30.05\%\\
Java	        & 39&20.21\%\\
C\#	            & 33&17.10\%\\
Compiled Python	& 30&15.54\%\\
Objective-C	    &15& 7.77\%	\\
Other	        &11& 5.70\%	\\
Swift	        &7& 3.63\%	\\ \hline

\end{tabular}
\end{center}
\label{table:c4}

\begin{center}
\caption*{Table C.5. Software/Systems encountered when reverse engineering.}
\begin{tabular}{p{.64\NetTableWidth} p{.12\NetTableWidth} p{.24\NetTableWidth}}
\hline
\textbf{Software/System} & \textbf{Count} & \textbf{Percentage} \\ \hline
Malware	                        & 50 & 28.41\%	\\
Desktop applications 	        & 40 & 22.73\%	\\
Mobile applications	            & 25 & 14.20\%	\\
Kernel	                        & 22 & 12.50\%	\\
Embedded systems including IoT	& 17 & 9.66\%	\\
Firmware	               &     16  & 9.09\%	\\
Other	                      &  6 &  3.41\%	\\\hline
 \end{tabular}

\end{center}
\end{table}
\label{table:c5}

%% file: sections/appendix_d.tex
\clearpage

\twocolumn[\section*{\hfil Appendix D. Ghidra v.s. IDA\hfil}]

\label{sec:appendixd}

\begin{table}[!hbtp]
\begin{center}
\caption*{Table D.1. Free or commercial edition of IDA.}
\begin{tabular}{p{.64\NetTableWidth} p{.12\NetTableWidth} p{.24\NetTableWidth}}
\hline
 \textbf{IDA Edition} & \textbf{Count} & \textbf{Percentage} \\ \hline
Free	    & 25 & 50\% \\
Commercial	& 25 & 50\%	\\\hline
 \end{tabular}
\end{center}
\label{table:d1}

\begin{center}
\caption*{Table D.2. IDA/IDA Pro GUI features that should be improved.}
\begin{tabular}{p{.64\NetTableWidth} p{.12\NetTableWidth} p{.24\NetTableWidth}}
\hline
 \textbf{Feature} & \textbf{Count} & \textbf{Percentage} \\ \hline 
Debugger	                            & 18 & 30\%	\\
Third-party plugins	                    & 11 & 18.33\%	\\  
Disassembler	                        & 9 & 15\%	\\
Graph view	                            & 6 & 10\% \\
Function call graph	                    & 4 & 6.67\% \\	
Library function signature recognition	& 4 & 6.67\% \\		
Reference table	                        & 3 & 5\% \\
Hex view	                            & 2 & 3.33\% \\	
Variable renaming	                    & 0 & 0.00\% \\
Function signature modification	        & 0 & 0.00\% \\	
Other	                                & 3 & 5\% \\	\hline
 \end{tabular}
\end{center}
\vspace{-0.1in}
\label{table:d2}

\begin{center}
\caption*{Table D.3. Does IDA/IDA Pro have any features that are advantageous over those in Ghidra?}
\begin{tabular}{p{.64\NetTableWidth} p{.12\NetTableWidth} p{.24\NetTableWidth}}
\hline
 \textbf{} & \textbf{Count} & \textbf{Percentage} \\ \hline
Yes	& 13 & 48.15\%	\\ 
No	& 14 & 51.85\%	\\ \hline

 \end{tabular}
\end{center}
\vspace{-0.1in}
\label{table:d3}

\caption*{Table D.4. Ghidra GUI features that should be improved.}
\begin{center}
\begin{tabular}{p{.64\NetTableWidth} p{.12\NetTableWidth} p{.24\NetTableWidth}}
\hline
 \textbf{Feature} & \textbf{Count} & \textbf{Percentage} \\ \hline
Debugger	                            & 9 & 20.93\%	\\
Graph view	                            & 8 & 18.60\% \\	
Library function signature recognition	& 8 & 18.60\% \\
Function call graph	                    & 6 & 13.95\% \\	
Hex view	                            & 6 & 13.95\% \\	
Disassembler	                        & 3 & 6.98\%	\\  
Function signature modification	        & 1 & 2.33\% \\
Third-party plugins	                    & 1 & 2.33\%	\\	
Data type viewer                        & 1 & 2.33\% \\
Reference table	                        & 0 & 0.00\% \\	\hline
 \end{tabular}
\end{center}
\vspace{-0.1in}
\label{table:d4}

\begin{center}
\caption*{Table D.5. Does Ghidra have any features which are better than those in IDA/IDA Pro?}
\begin{tabular}{p{.64\NetTableWidth} p{.12\NetTableWidth} p{.24\NetTableWidth}}
\hline
 \textbf{} & \textbf{Count} & \textbf{Percentage} \\ \hline
Yes	& 13 & 52.00\%	\\ 
No	& 12 & 48.00\%	\\ \hline
 \end{tabular}
\end{center}
\end{table}
\label{table:d5}

%% file: sections/appendix_e.tex
\clearpage
\twocolumn[\section*{\hfil Appendix E. Teaching Reverse Engineering\hfil}]

\label{sec:appendixe}

\begin{table}[!ht]
\begin{center}
\caption*{Table E.1. Do you teach or have you ever taught reverse engineering.}
\begin{tabular}{p{.64\NetTableWidth} p{.12\NetTableWidth} p{.24\NetTableWidth}}
\hline
 \textbf{} & \textbf{Count} & \textbf{Percentage} \\ \hline
No	    & 48 & 66.67\% \\ 	
Yes	    & 24 & 33.33\% \\ \hline
 \end{tabular}
\end{center}
\end{table}
\label{table:e1}

\begin{table}[!ht]
\begin{center}
\caption*{Table E.2. Resources used to teach reverse engineering.}
\begin{tabular}{p{.64\NetTableWidth} p{.12\NetTableWidth} p{.24\NetTableWidth}}
\hline
 \textbf{Resource} & \textbf{Count} & \textbf{Percentage} \\ \hline
Lab activities	& 20 & 27.28\% \\ 
Demonstration	& 19 & 26.39\% \\ 
Textbooks	    & 11  & 15.28\% \\ 	
Websites	    & 11  & 15.28\% \\ 	
Online videos	& 10  & 13.89\% \\ 
Other	        & 1 & 1.39\% \\ \hline
 \end{tabular}
\end{center}
\end{table}
\label{table:e2}

\begin{table}[!ht]
\begin{center}
\caption*{Table E.3. Types of assignments used to teach students reverse engineering.}
\begin{tabular}{p{.64\NetTableWidth} p{.12\NetTableWidth} p{.24\NetTableWidth}}
\hline
 \textbf{Assignment Type} & \textbf{Count} & \textbf{Percentage} \\ \hline
Projects & 15 & 37.50\%	\\ 
CTF-style Assignments & 15 & 37.50\%\\
Exams	 & 5 & 12.50\% \\ 
Written assignments	& 3 & 7.50\%\\  
Other	& 2 & 5.00\%	\\
- Real Malware && \\
- Labs &&\\ \hline
 \end{tabular}
\end{center}
\end{table}
\label{table:e3}

\begin{table}[!ht]
\begin{center}
\caption*{Table E.4. Methods used to assess student's knowledge of reverse engineering.}
\begin{tabular}{p{.64\NetTableWidth} p{.12\NetTableWidth} p{.24\NetTableWidth}}
\hline
 \textbf{Method} & \textbf{Count} & \textbf{Percentage} \\ \hline
In-class Presentations	& 8 & 25.00\%\\ 
Exams	& 7 & 21.88\% \\ 
Timed assignments	& 6 & 18.75\%\\ 
Informal Verbal Assessment	& 4 & 12.50\%\\ 
Other	& 7 & 21.88\%\\
- Projects & &\\
- Labs & &\\
\hline
 \end{tabular}
\end{center}
\end{table}
\label{table:e4}

\begin{table}[!ht]
\begin{center}
\caption*{Table E.5. Challenges in teaching reverse engineering.}
\begin{tabular}{p{.64\NetTableWidth} p{.12\NetTableWidth} p{.24\NetTableWidth}}
\hline
 \textbf{Challenge} & \textbf{Count} & \textbf{Percentage} \\ \hline
Students' lack of prerequisite knowledge	    & 18 & 29.51\%	\\ 
Students' understanding of assembly language    & 13 & 21.31\%	\\ 
Learning curve of reverse engineering software	& 11 & 18.03\%	\\ 
Lack of reverse engineering educational resources & 8 &	13.11\%	\\
Quality of existing reverse engineering educational resources & 7 &	11.48\% \\
Usability of reverse engineering software	    & 4 & 6.56\%	\\  \hline
 \end{tabular}
\end{center}
\end{table}
\label{table:e5}

\begin{table}[!ht]
\begin{center}
\caption*{Table E.6. Tools used in classes to teach reverse engineering.}
\begin{tabular}{p{.64\NetTableWidth} p{.12\NetTableWidth} p{.24\NetTableWidth}}
\hline
 \textbf{Tool} & \textbf{Count} & \textbf{Percentage} \\ \hline
IDA/IDA Pro	       & 14 & 14.29\%	\\ 
Ghidra	           & 10 & 10.20\%	\\ 
GDB	               & 10 & 10.20\%	\\ 
Objdump	           & 10 & 10.20\%	\\ 
Radare2	           & 7 & 7.14\%	\\ 
Other	           & 6 & 6.12\%	\\
CFF Explorer	   & 6 & 6.12\%	\\ 
YARA	           & 4 & 4.08\%	\\ 
Cuckoo Sandbox	   & 4 & 4.08\%	\\ 
Binwalk	           & 4 & 4.08\%	\\ 
ApkTool	           & 4 & 4.08\%	\\ 
Remnux	           & 3 & 3.06\%	\\ 
Peda	           & 3 & 3.06\%	\\ 
Hopper	           & 3 & 3.06\%	\\ 
Dex2jar	           & 3 & 3.06\%	\\ 
Frida	           & 2 & 2.04\%	\\ 
Jeb	               & 1 & 1.02\%	\\ 
Jadx               & 1 & 1.02\%	\\ 
Binary Ninja	   & 1 & 1.02\%	\\ 
Angr	           & 1 & 1.02\%	\\ 
Androguard	       & 1 & 1.02\%	\\ 
Uncompyle	       & 0 & 0.00\%	\\ 
Triton	           & 0 & 0.00\%	\\ 
Qira	           & 0 & 0.00\%	\\ 
Objection          & 0 & 0.00\%	\\ 
Karakatau          & 0 & 0.00\%	\\ 
Cheat Engine	   & 0 & 0.00\%	\\ 
Boomerang	       & 0 & 0.00\%	\\ 
Barf	           & 0 & 0.00\%	\\ 
Apk2Gold	       & 0 & 0.00\%	\\ 
 \hline
 \end{tabular}
\end{center}
\end{table}
\label{table:e6}\clearpage

%% file: sections/appendix_f.tex
\twocolumn[\section*{\hfil Appendix F. Web-Based Reverse Engineering\hfil}]

\begin{table}[!ht]

\label{sec:appendixf}
\begin{center}
\caption*{Table F.1. Do you know of a web-based reverse engineering platform.}
\begin{tabular}{p{.64\NetTableWidth} p{.12\NetTableWidth} p{.24\NetTableWidth}}
\hline
 \textbf{} & \textbf{Count} & \textbf{Percentage} \\ \hline
No &49&77.78\%\\
Yes	&14&22.22\%\\
\hline
\end{tabular}
\end{center}
\label{table:f1}

\begin{center}
\caption*{Table F.2. Interested in a web-based reverse engineering platform?}
\begin{tabular}{p{.64\NetTableWidth} p{.12\NetTableWidth} p{.24\NetTableWidth}}
\hline
 \textbf{} & \textbf{Count} & \textbf{Percentage} \\ \hline
Maybe	&31&47.69\%\\
Yes	    &24&36.92\%\\
No	    &10&15.38\% \\
\hline
 \end{tabular}
\end{center}
\label{table:f2}

\begin{center}
\caption*{Table F.3. Desired characteristics or aspects of a web-based reverse engineering platform.}
\begin{tabular}{p{.64\NetTableWidth} p{.12\NetTableWidth} p{.24\NetTableWidth}}
\hline
 \textbf{Characteristic} & \textbf{Count} & \textbf{Percentage} \\ \hline
Access to various tools/features	&46&21.30\%\\
Allows for collaboration with other users	&40&18.52\%\\
Ability to create educational lessons	&31&14.35\%\\
CTF style challenges	&30&13.89\%\\
Community forums	&20&9.26\%\\
Annual competition	&15&6.94\%\\
Ability to generate groups	&13&6.02\%\\
Online job fair opportunities	&13&6.02\%\\
User ranking system	&8&3.70\%\\
\hline
 \end{tabular}
\end{center}
\label{table:f3}
\end{table}
\clearpage